\documentclass[aps,prd,reprint,groupedaddress]{revtex4-2}

\usepackage[T1]{fontenc}
\usepackage[utf8]{inputenc}
\usepackage{amsfonts,amsmath,amssymb,accents,mathrsfs}
\usepackage[retainorgcmds]{IEEEtrantools}
\usepackage{youngtab}
\usepackage{microtype}
\usepackage[colorlinks,
			breaklinks,
			hyperfootnotes,
			linktoc=page,
            linkcolor=blue,
            citecolor=blue,
            urlcolor=blue,
            unicode,
            pdfstartview=FitH,
            bookmarks,
            bookmarksnumbered]{hyperref}

\renewcommand\a {{\alpha}}
\renewcommand\b {{\beta}}
\newcommand\g {{\gamma}}
\renewcommand\d {{\delta}}
\renewcommand\r {{\rho}}

\renewcommand\L {{\Lambda}}

\newcommand\ad {{\dot{\alpha}}}
\newcommand\bd {{\dot{\beta}}}
\newcommand\gd {{\dot{\gamma}}}

\newcommand\Ysf{{\textsf{Y}}}
\newcommand\N{{\mathcal{N}}}
\newcommand\E{{\mathcal{E}}}
\newcommand\I{{\mathcal{I}}}
\renewcommand\H{{\mathcal{H}}}
\newcommand\W{{\mathcal{W}}}
\newcommand\K{{\mathcal{K}}}
\newcommand\V{{\mathcal{V}}}
\newcommand\J{{\mathcal{J}}}
\renewcommand\O{{\varOmega}}

\newcommand\D {{\rm D}}
\newcommand\Dd {{\bar{\rm D}}}
\newcommand\pa {{\partial}}

\def\n{\IEEEyesnumber}
\def\sn{\IEEEyessubnumber}

\begin{document}


\title{Superspace First-order Formalism for Massless Arbitrary Superspin Supermultiplets}


\author{Konstantinos~Koutrolikos}
\email{konstantinos\_koutrolikos@brown.edu}
\affiliation{Brown Theoretical Physics Center}
\affiliation{Department of Physics, Brown University}



\begin{abstract}
A new description of free massless superfields of arbitrary superspin $\Ysf$ ($\Ysf>1/2$) is proposed.
Following the first-order philosophy, we relax some of the properties (reality, gauge redundancy) of the
unconstrained higher spin prepotentials and we construct first and half order invariants quantities. These are
used to write trivially invariant actions. Additional auxiliary superfields that play the role of spin
connections are used to enforce a new local symmetry that restores the degrees of freedom.
\end{abstract}


\maketitle

\section{Introduction}
The general Ostrogradskiy procedure reduces the order of derivatives in the Lagrangian (or Hamiltonian)
description of a theory by introducing new variables.  A specific application of this procedure is known as
first order formalism.  The power of this method has been epitomized in the frame formulation of gravity and
supergravity.  Not only first-order formalism is necessary for the gauge theory approach to gravitation but it
is also recommended for the path integral quantization. By using the frame field and not the metric in the
path integral, one avoids the complication of preserving the correct spacetime signature and additionally the
gravitational and matter Lagrangians take a polynomial form which is crucial for non-perturbative effects.

For higher spin theories, first order formalism is particularly useful. For a spin $s$ gauge field the
simplest gauge invariant object, \emph{field strength}, carries $s$ derivatives and thus can not be used for
writing a two derivative action. In the metric-like description of higher spin gauge fields
\cite{Fronsdal:1978rb,Fang:1978wz}, this problem is addressed by introducing a tower of higher spin
connections \cite{deWit:1979sib} and identify appropriate reductions of the gauge symmetry group that allow
the construction of invariant quantities with two derivatives ---Frosdal's equations of motion.
Alternatively, one can use the first order formalism. By introducing additional auxiliary degrees of freedom
it is possible to construct invariant quantities with only one derivative which can be used to write the two
derivative higher spin Lagrangian as a second order polynomial and make it manifestly gauge invariant. This
first-order description of higher spin gauge fields was developed by Vasiliev
\cite{Vasiliev:1980as,Vasiliev:1986td}.

The metric-like description offers a more geometric viewpoint to higher spin theory
that extends our spin 2 intuition, provides a very economic description in terms of the number of fields it
requires and has been used to construct various consistent higher spin interactions
\cite{Berends:1985xx,Barnich:1993vg,Bekaert:2009ud,Bekaert:2010hk,Buchbinder:2012iz,Joung:2012fv,Joung:2013nma}.
On the other hand the frame-like description generalizes the gauge approach to higher spins, provides an
economy of ideas that underlie YM, GR and higher spin, and it has been the most successful approach towards
constructing consistent interactions among higher spins
\cite{Fradkin:1987ks,Lopatin:1987hz,Vasiliev:1987tk,
    Vasiliev:2001wa,Alkalaev:2002rq,Zinoviev:2008ze,Ponomarev:2010st,Zinoviev:2010cr,
    Boulanger:2011qt,Zinoviev:2011fv,Buchbinder:2019dof,Buchbinder:2019olk,Buchbinder:2019kuh,
Buchbinder:2020rex,Khabarov:2020bgr,Khabarov:2021xts}.

Of course, understanding interactions involving higher spins is a necessary condition for understanding string
theory as string interactions allow the exchange of higher spin states. From this perspective one should
consider supersymmetric higher spin theories as for most formulations of string theory supersymmetry is a
necessary ingredient. Manifestly supersymmetric theories of higher spins have been constructed using the superspace
analog of the metric-like description
\cite{Kuzenko:1993jq,Kuzenko:1993jp,Gates:2013rka,Gates:2013ska,Buchbinder:2020yip,Koutrolikos:2020tel,Buchbinder:2021ite}
and various cubic interactions of them with matter supermultiplets have been found
\cite{Buchbinder:2017nuc,Hutomo:2017phh,Hutomo:2017nce,Koutrolikos:2017qkx,
Buchbinder:2018wwg,Buchbinder:2018nkp,Buchbinder:2018gle,Hutomo:2018tjh,Buchbinder:2022kzl}. Moreover, cubic
interactions among higher spin supermultiplets of arbitrary superspin $\Ysf$ and supermultiplets with
half-integer [$(s+1/2)-\Ysf-\Ysf$] or integer superspins [$s-\Ysf-\Ysf$]  have also been constructed
\cite{Buchbinder:2018wzq,Gates:2019cnl}. These interactions are of the abelian type because the cubic
superspace Lagrangian is of the form $\mathcal{L}_1\sim\Phi_1~W_2~W_3$, where $\Phi_1$ is the set of
superfields that describe the superspin $\Ysf_1$ and $W_2, W_3$ are the gauge
invariant superfield strengths for superspins $\Ysf_2,\Ysf_3$ respectively.

It would be desirable to consider cubic interactions of the non-abelian type
$\mathcal{L}_1\sim\Phi_1~\Phi_2~W_3$.  This class of interactions are in general more interesting because they
have the potential to generate non-trivial deformations of the gauge symmetry that may also deform the gauge
algebra.  Such cubic interactions have been recently constructed for higher spin theories with on-shell
supersymmetry \cite{Khabarov:2020deh}. An important subclass of such interactions are the electromagnetic
interactions of higher spin multiplets, [$\Ysf-\Ysf-1/2$] where the two higher spin supermultiplets form a
doublet under U(1) ---must come in pairs of opposite charges--- and couple to the vector supermultiplet. In
\cite{Buchbinder:2021igw} such non-abelian interaction was constructed for the non-minimal supergravity
supermultiplet [$3/2-3/2-1/2$]. This was achieved by developing a first order description of the non-minimal
supergravity  supermultiplet in superspace.

In conventional superspace description of supergravity the superframes are constrained superfields in order to
eliminate the extra degrees of freedom they carry. Solving these constraints while maintaining supersymmetry
manifest is possible by expressing the superframe in terms of a set of unconstrained prepotential superfields,
which include a complex vector superfield $\H_{\a\ad}$. This solution also introduces new redundancies for the
prepotentials in addition to superspace general covariance and superlocal Lorentz rotations. All these
symmetries can be used to eliminate all prepotentials except the real part of the vector prepotential
$H_{\a\ad}$.  In \cite{Buchbinder:2021igw} ---while attempting to streamline the construction of non-abelian
cubic interactions of the supergravity supermultiplet--- the complex nature of the vector prepotential was
restored, $(H_{\a\ad}\to\H_{\a\ad})$ while at the same time a new local symmetry
$(\d_\eta\H_{\a\ad}=i\eta_{\a\ad},~\eta_{\a\ad}=\bar{\eta}_{\a\ad})$ was introduced in order to
remove the added auxiliary degrees of freedom. This symmetry corresponds to the linearized coordinate
transformation of the superframe's vector prepotential.

This approach naturally leads to the development of a first-order formalism for linearized supergravity
without using the conventional constrained superframes but instead it utilizes the unconstrained superfields
of free theory. Following the first order philosophy:
(\emph{i}) We introduced new degrees of freedom by complexifying the linearized supergravity superfield
\begin{equation}
    H_{\a\ad}~\to~\mathcal{H}_{\a\ad}
\end{equation}
(\emph{ii}) The relaxed gauge transformation of $\mathcal{H}_{\a\ad}$ allowed us to define a simpler gauge
invariant quantities $\I_{\b\a\ad}$
\begin{equation}
    \I_{\b\a\ad}=\D_{\b}\mathcal{H}_{\a\ad}+C_{\b\a}\bar{\chi}_{\ad}
\end{equation}
where $\chi_{\a}$ is the compensating superfield.\\
(\emph{iii}) $\I_{\b\a\ad}$ satisfies various identities which can be interpreted as Bianchi identities for
additional symmetries.  These symmetries correspond to the symmetries discussed above and are implemented in
the action by the introduction of a pair of auxiliary, connection-like superfields
$\O_{\b\a\ad},~\W_{\b\a\ad}$
\begin{equation}
S\hspace{-1.0mm}=\hspace{-1.8mm}\int\hspace{-1.5mm} d^8z\Big\{
\W^{\b\a\ad}\O_{\b\a\ad}+\W^{\b\a\ad}\I_{\b\a\ad}+\O^{\b\a\ad}\J_{\b\a\ad}\Big\}+c.c.
\end{equation}
where $\J_{\b\a\ad}$ depends only on derivatives of $\I_{\b\a\ad}$.\\
(\emph{iv}) One of these additional symmetries is local and algebraic in nature. Hence it is used to eliminate
the extra degrees introduces in (\emph{i}). After integrating out $\O_{\b\a\ad}$ and $\W_{\b\a\ad}$, we
recover the linearized supergravity action described by the prepotential superfield
$H_{\a\ad}=\H_{\a\ad}+\bar{\H}_{\a\ad}$ and the compensator $\chi_{\a}$.

One approach towards constructing consistent interactions of higher superspins in superspace would be to
generalize the conventional supergravity description by considering higher rank superframes and
superconnections.  Such superfields, like supergravity, will carry too many degrees of freedom and must be
constrained in order to describe just the irreducible higher superspins. However, unlike supergravity, it is
not currently known if such set of constraints exist and how to determine them. An idea is to use the higher
rank superframes and superconnections in combination with higher rank symmetry generators to define
generalized super-covariant derivatives. Their superalgebra will form a supersymmetric higher spin algebra and
can be used to define generalized supertorsions and supercurvatures.  The hope is that there is a set of
constraints for the generalized supertorsions and supercurvatures which are compatible with supersymmetry
algebra, the super-Jacobi identities and describe higher superspin irreducible representations.

A different approach is to bypass all the above and consider the unconstrained prepotential
superfields \footnote{These higher spin prepotentials should emerge as the superfield solution of the
constraints for the higher rank superframes} used in the formulation of the free higher superspin theory
---which is the starting point for cubic interactions--- and develop a first order formalism similar
to the one developed in \cite{Buchbinder:2021igw} for the supergravity supermultiplet and described above.

In this work, we show that indeed such a first order formulation of all irreducible $4D,\mathcal{N}=1$
higher spin supermultiplets exist. We find that for half integer superspin supermultiplets $(s+1,s+1/2)$
there are two first order descriptions which correspond to the minimal and non-minimal descriptions of the
supermultiplet. For integer superspins $(s+1/2,s)$ ($s>1$) there is a unique first order description.

\section{Half integer superspin}
The half-integer superspin supermultiplet $\Ysf=s+1/2$ on-shell describes the propagation of
massless $j=s+1$ and $j=s+1/2$ spins. The superspace realization of this supermultiplet is given by the
equivalence class $[H_{\a(s)\ad(s)}]$ of an independently symmetric, real  $(s,s)$ SL(2,$\mathbb{C}$)
superfield  tensor
$H_{\a(s)\ad(s)}$~\footnote{We use the conventions of \emph{Superspace}\cite{Gates:1983nr}. The notation
$\a(s)$ means that there are $s$ undotted spinorial indices $\a_1\a_2\dots\a_{s}$ which are symmetrized.
Similarly for $\ad(s)$}
defined by the
equivalence relation (for $s>0$):
\begin{IEEEeqnarray*}{ll}
    \hat{H}_{\a(s)\ad(s)}~\sim~H_{\a(s)\ad(s)}&~+\tfrac{1}{s!}~\D_{(\a_s}\bar{L}_{\a(s-1))\ad(s)}\n\\
                                              &~-\tfrac{1}{s!}~\Dd_{(\ad_s}L_{\a(s)\ad(s-1))}
\end{IEEEeqnarray*}
This redundancy was initially postulated in \cite{Kuzenko:1993jp} and it was later shown in
\cite{Gates:2013rka}
to be a consequence of demanding a smooth transition between the Lagrangian description of
massive half-integer superspins \cite{Koutrolikos:2020tel} and the Lagrangian description of massless
half-integer superspins.  The simplest gauge invariant, \emph{superfield strength}, is
\begin{equation}
    W_{\a(2s+1)}=\Dd^2\D_{(\a_{2s+1}}\pa_{\a_{2s}}{}^{\ad_{1}}\ldots\pa_{\a_{s+1}}{}^{\ad_s}H_{\a(s))\ad(s)}
\end{equation}
and was constructed first in \cite{Gates:1983nr}.

Following the procedure in \cite{Buchbinder:2021igw}, we
introduce new degrees of freedom by complexifying the superfield $H_{\a(s)\ad(s)}$ and its $L$-transformation:
\begin{IEEEeqnarray*}{c}\n
H_{\a(s)\ad(s)}~\to~\H_{\a(s)\ad(s)}\sn\label{cH}\\[2pt]
\d_{L}\H_{\a(s)\ad(s)}=\tfrac{1}{s!}\D_{(\a_s}\bar{L}_{\a(s-1))\ad(s)}\sn\label{dL}
\end{IEEEeqnarray*}
The effect of this on the higher spin fields generated by $\H_{\a(s)\ad(s)}$ is to relax the symmetrization of
their spacetime indices and introduce non-symmetric fields, in a manner analogous to \cite{Vasiliev:1980as}.

First order formalism, being a special case of the Ostrogradskiy procedure, is based on the factorization of
the operators $\Box=\pa^m\pa_m$ and $\pa_m\pa^n$ that appears in the second order Fronsdal equation of motion.
In superspace, the corresponding operators are
$\delta_{\ad}{}^{\gd}\D^{\b}\Dd^2\D_{\b}$=$-\D^{\b}\Dd_{\ad}\Dd^{\gd}\D_{\b}$,
$\delta_{\a}{}^{\g}\Dd_{\ad}\D^2\Dd^{\bd}$=$-\D_{\ad}\D_{\a}\D^{\g}\Dd^{\gd}$ and
$\D_{\a}\Dd_{\ad}\D^{\g}\Dd^{\gd}$. Based on these factorizations,
we attempt to construct a first-order invariant for $\H_{\a(s)\ad(s)}$ by considering the following:
\begin{IEEEeqnarray*}{ll}
    I_{\b\a(s)\ad(s-1)}=&~\Dd^{\ad_s}\D_{\b}\H_{\a(s)\ad(s)}\n\\
                        &+\frac{A}{s!}~C_{\b(\a_s}~\D^{\g}\Dd^{\gd}\H_{|\g|\a(s-1))\gd\ad(s-1)}~.
\end{IEEEeqnarray*}
The transformation of $I_{\b\a(s)\ad(s-1)}$ under \eqref{dL} is
\begin{IEEEeqnarray*}{l}
\d_{L}I_{\b\a(s)\ad(s-1)}=\n\\[2pt]
\hfill-\frac{A+1}{s!}C_{\b(\a_s}\Dd^{\ad_{s}}\Big[\D^2\bar{L}_{\a(s-1))\ad(s)}
                        -\Dd^{\ad_{s+1}}\bar{\L}_{\a(s-1))\ad(s+1)}\Big]\\[2pt]
   +\frac{A}{s!}~C_{\b(\a_s}\D_{\a_{s-1}}\Big[
\D^{\g}\Dd^{\gd}\bar{L}_{\g\a(s-2))\gd\ad(s-1)}\\
\hspace{28mm}+\frac{s-1}{s}\Dd^{\gd}\D^{\g}\bar{L}_{\g\a(s-2))\gd\ad(s-1)}\\[2pt]
                  \hspace{28mm}-\frac{1}{(s-2)!}~\D_{(\a_{s-2}}\bar{\L}_{\a(s-3)))\ad(s-1)}\Big]
\end{IEEEeqnarray*}
where the two extra parameters $\bar{\L}_{\a(s-1)\ad(s+1)}$ and $\bar{\L}_{\a(s-3)\ad(s-1)}$ correspond
to additional symmetries emerging due to the anticommuting nature of the spinorial covariant derivatives.
If we insist on demanding a first-order invariant quantity, then it becomes obvious from the above that
we must consider two cases. Either choose $A=0$ and introduce a fermionic compensating superfield
$\chi_{\a(s)\ad(s-1)}$ with transformation
$\d\chi_{\a(s)\ad(s-1)}$=$\Dd^2L_{\a(s)\ad(s-1)}$+$\D^{\a_{s+1}}\L_{\a(s+1)\ad(s-1)}$ or $A$=$-1$
and consider a different compensating superfield $\chi_{\a(s-1)\ad(s-2)}$ with
$\d\chi_{\a(s-1)\ad(s-2)}$
=$\Dd^{\ad_{s-1}}\D^{\a_{s}}L_{\a(s)\ad(s-1)}$ +
$\frac{s-1}{s}$$\D^{\a_{s}}\Dd^{\ad_{s-1}}L_{\a(s)\ad(s-1)}$
+$\frac{1}{(s-2)!}$$\Dd_{(\ad_{s-2}}\L_{\a(s-1)\ad(s-3))}$.

The two cases will correspond respectively to the non-minimal and minimal formulation of half-integer
superspins as described in \cite{Gates:2013rka}. It is very satisfying that just the requirement of a
first order description determines all different
\emph{variant} formulations of the theory and generates all required superfields for each one of
them. The two $L$-invariant building blocks are:
\begin{IEEEeqnarray*}{ll}
A=0:&\I_{\b\a(s)\ad(s-1)}=\Dd^{\ad_{s}}\I_{\b\a(s)\ad(s)},\n\label{nmI}\\[1pt]
     &\I_{\b\a(s)\ad(s)}=\D_{\b}\H_{\a(s)\ad(s)}+\frac{1}{s!}C_{\b(\a_s}\bar{\chi}_{\a(s-1))\ad(s)}\\[2pt]
A=-1:&~\I_{\b\a(s)\ad(s-1)}=\Dd^{\ad_{s}}\D_\b\H_{\a(s)\ad(s)}\n\label{mI}\\[1pt]
      &\hspace{22mm}-\frac{1}{s!}C_{\b(\a_{s}}\D^{\g}\Dd^{\ad_s}\H_{|\g|\a(s-1))\ad(s)}\\[1pt]
      &\hspace{22mm}-\frac{1}{s!}C_{\b(\a_s}\D_{\a_{s-1}}\bar{\chi}_{\a(s-2))\ad(s-1)}
    \end{IEEEeqnarray*}
It is interesting to observe that for supersymmetric theories Ostrogradskiy's procedure does not stop at
first order operators. Because
of the supersymmetry algebra some of the first order operators can be further factorized and one can construct
\emph{$1/2$-order} invariants. An example is
\begin{equation}\label{mK}
    \K_{\a(s+1)\ad(s)}=\frac{1}{(s+1)!}~\D_{(\a_{s+1}}\H_{\a(s))\ad(s)}
\end{equation}
which is invariant under \eqref{dL}. For the $A=0$ case \eqref{nmI}, $\K_{\a(s+1)\ad(s)}$ is not an
independent invariant quantity as it is captured by the symmetric part of $\I_{\b\a(s)\ad(s)}$ which is also
an $1/2$-order $L$-invariant. On the other hand, for the $A=-1$ case \eqref{mI} $K_{\a(s+1)\ad(s)}$ is a new
and independent $L$-invariant quantity which must be used together with $\I_{\b\a(s)\ad(s-1)}$ in order to
derive the first order action.

It becomes evident that the difference between the two formulations of $4D,~\N=1$
half-integer superspins is that there is a fundamental half-order invariant in one case and all higher order
invariants are generated by its derivatives. In the other case, there are two independent invariants of
half and first order respectively.
\subsection{First order formalism of non-minimal half-integer superspin supermultiplets}
The advantage of having a half or first order $L$-invariant building blog as $\I_{\b\a(s)\ad(s)}$
\begin{equation}
    \I_{\b\a(s)\ad(s)}=\D_{\b}\H_{\a(s)\ad(s)}+\frac{1}{s!}~C_{\b(\a_s}~\bar{\chi}_{\a(s-1))\ad(s)}\label{I1}
\end{equation}
is that the we can write actions $S$=$S[\I]$ which are trivially $L$-invariant.
However, there are a couple more symmetries that we want to impose. The first one is the $\L$ redundancy in
the definition of the compensator $\chi_{\a(s)\ad(s-1)}$
\begin{IEEEeqnarray*}{l}
\d_{\L}\chi_{\a(s)\ad(s-1)}=\D^{\a_{s+1}}\L_{\a(s+1)\ad(s-1)}\n\\[1pt]
\d_{\L}\I_{\b\a(s)\ad(s)}=-\frac{1}{s!}~C_{\b(\a_s}~\Dd^{\ad_{s+1}}\bar{\L}_{\a(s-1))\ad(s+1)}\n\label{dLambdaI1}
\end{IEEEeqnarray*}

The second is a new local symmetry that we have to impose in order to remove the extra degrees of freedom
introduced in the theory via complexification \eqref{cH}:
\begin{IEEEeqnarray*}{l}
    \d_\eta\H_{\a(s)\ad(s)}=i~\eta_{\a(s)\ad(s)}~,~\eta_{\a(s)\ad(s)}=\bar{\eta}_{\a(s)\ad(s)}\n\label{detaH}\\[1pt]
    \d_\eta\I_{\b\a(s)\ad(sr)}=i~\D_{\b}\eta_{\a(s)\ad(s)}\n\label{detaI1}~.
\end{IEEEeqnarray*}
This is a direct generalization to higher spins of the transformation introduced in
\cite{Buchbinder:2021igw} which corresponds to a change of coordinates transformation of superframe's
linearized vector prepotential.

The above symmetries will be implemented by an appropriate set of auxiliary superfields which play the role of
generalized superconnections. Because $\I_{\b\a(s)\ad(s)}$ is a half-order invariant in order to write an
action we require a pair of them $(\W_{\b\a(s)\ad(s)}, \O_{\b\a(s)\ad(s)})$
\begin{IEEEeqnarray*}{ll}
S=\int d^8z\Big\{
&\W^{\b\a(s)\ad(s)}\O_{\b\a(s)\ad(s)}+\W^{\b\a(s)\ad(s)}\I_{\b\a(s)\ad(s)}\\
&+\O^{\b\a(s)\ad(s)}\J_{\b\a(s)\ad(s)}\Big\}+c.c.\n\label{S1}
\end{IEEEeqnarray*}
where $\J_{\b\a(s)\ad(s)}$ is a $3/2$-order invariant
---not to be confused with the $1.5$-order formalism of (super)gravity---
that can be expressed purely in terms of derivatives of
$\I_{\b\a(s)\ad(s)}$
\begin{widetext}
\begin{IEEEeqnarray*}{ll}
    \J_{\b\a(s)\ad(s)}=&~f_1~\D^2\I_{\b\a(s)\ad(s)}+f_2~\Dd^2\I_{\b\a(s)\ad(s)}
                       +f_3~\D_{\b}\Dd^\bd\bar{\I}_{\a(s)\bd\ad(s)}
                       +f_4~\Dd^\bd\D_{\b}\bar{\I}_{\a(s)\bd\ad(s)}\n\\
                       &+\frac{g_1}{s!}~C_{\b(\a_s}\Dd^2\I_{\a(s-1))\ad(s)}
                       +\frac{g_2}{s!}~\D_\b\Dd_{(\ad_s}\bar{\I}_{\a(s)\ad(s-1))}
                       +\frac{g_3}{s!}~\Dd_{(\ad_s}\D_\b\bar{\I}_{\a(s)\ad(s-1))}\\
                       &+\frac{d_1}{s!s!}~C_{\b(\a_s}~\D^{\g}\Dd_{(\ad_s}\bar{\I}_{|\g|\a(s-1))\ad(s-1))}
                       +\frac{d_2}{s!s!}~C_{\b(\a_s}~\Dd_{(\ad_s}\D^{\g}\bar{\I}_{|\g|\a(s-1))\ad(s-1))}
\end{IEEEeqnarray*}
\end{widetext}
and $\I_{\a(s-1)\ad(s)}=C^{\b\a_s}\I_{\b\a(s)\ad(s)}$.
Checking the invariance of \eqref{S1} under symmetries \eqref{dLambdaI1} and \eqref{detaI1} is
a little more involved. This is the usual trade of first order formalism.
The process is simplified by assigning
\begin{IEEEeqnarray*}{ll}
    \d S\hspace{-1mm}=\hspace{-2mm}\int\hspace{-1mm} d^8z \Big\{&\d\W^{\b\a(s)\ad(s)}\I_{\b\a(s)\ad(s)}
                    +\d\O^{\b\a(s)\ad(s)}\J_{\b\a(s)\ad(s)}\\
                     &+\W^{\b\a(s)\ad(s)}\Big[\d\O_{\b\a(s)\ad(s)}+\d\I_{\b\a(s)\ad(s)}\Big]\n\label{dS1}\\
                     &+\O^{\b\a(s)\ad(s)}\Big[\d\W_{\b\a(s)\ad(s)}+\d\J_{\b\a(s)\ad(s)}\Big]\Big\}
                         \hspace{-1mm}+\hspace{-1mm}c.c.
\end{IEEEeqnarray*}
appropriate transformations to the auxiliary superfields.
We can eliminate the last two lines in the variation \eqref{dS1} by choosing the following:
\vspace*{-1mm}
\begin{IEEEeqnarray*}{l}
    \d\O_{\b\a(s)\ad(s)}=-\d\I_{\b\a(s)\ad(s)}\n\label{dO1}\\[3pt]
    \d_{L}\O_{\b\a(s)\ad(s)}=0~,\sn\\[1pt]
    \d_{\L}\O_{\b\a(s)\ad(s)}=\tfrac{1}{s!}C_{\b(\a_s}~\Dd^{\ad_{s+1}}\bar{\L}_{\a(s-1))\ad(s+1)}
    \sn\label{dLambdaO1}~,\\[1pt]
    \d_\eta\O_{\b\a(s)\ad(s)}=-i~\D_{\b}\eta_{\a(s)\ad(s)}~,\sn\label{detaO1}
    \end{IEEEeqnarray*}
\begin{widetext}
\vspace*{-4mm}
\begin{IEEEeqnarray*}{l}
    \d\W_{\b\a(s)\ad(s)}=-\d\J_{\b\a(s)\ad(s)}\n\label{dW1}\\[2pt]
    \d_{L}\W_{\b\a(s)\ad(s)}=0~,\sn\\[1pt]
\d_{\L}\W_{\b\a(s)\ad(s)}=\frac{f_1}{s!}~C_{\b(\a_s}\D^2\Dd^{\ad_{s+1}}\bar{\L}_{\a(s-1))\ad(s+1)}
+(f_3-\frac{s+1}{s}g_2)~\frac{1}{s!}\D_{\b}\Dd_{(\ad_s}\D^{\g}\L_{\g\a(s)\ad(s-1))}\sn\label{dLambdaW1}\\
\hspace{22mm}
-(f_4-\frac{s+1}{s}g_3)~\frac{1}{s1}\Dd_{(\ad_s}\D^2\L_{\b\a(s)\ad(s-1)}
-\frac{s+1}{s}\frac{d_1}{s!s!}~C_{\b(\a_{s}}\D^{\g}\Dd_{(\ad_s}\D^{\r}\L_{|\r\g|\a(s-1))\ad(s-1))}~,\\[2pt]
    \d_\eta\W_{\b\a(s)\ad(s)}=-i(f_2+f_4)\Dd^2\D_{\b}\eta_{\a(s)\ad(s)}
+i(2f_3-f_4-g_2)\D_{\b}\Dd^2\eta_{\a(s)\ad(s)}
+\frac{ig_1}{s!}~C_{\b(\a_s}\Dd^2\D^{\g}\eta_{|\g|\a(s-1))\ad(s)}\sn\label{detaW1}\\
\hspace{18mm}
-\frac{id_1}{s!}~C_{\b(\a_s}\D^{\g}\Dd^2\eta_{|\g|\a(s-1))\ad(s)}
+\frac{ig_3}{s!}~\Dd_{(\ad_s}\D_\b\Dd^{\gd}\eta_{\a(s)|\gd|\ad(s-1))}
+\frac{id_2}{s!s!}~C_{\b(\a_s}\Dd_{(\ad_s}\D^\g\Dd^{\gd}\eta_{|\g|\a(s-1))|\gd|\ad(s-1))}.
    \end{IEEEeqnarray*}
\end{widetext}
Using equations \eqref{dS1}, \eqref{detaO1} and \eqref{detaW1} we find that the $\eta$-invariance of the
action $(\d_{\eta}S=0)$ requires:
\begin{IEEEeqnarray*}{l}
    f_2=-f_4~,~g_1=0~,~g_2=4f_3-2f_4\n\label{eta1}\\[2pt]
    g_3=0~,~d_1=4f_3-2f_4~,~d_2=0
\end{IEEEeqnarray*}
For $\L$-invariance we substitute \eqref{dLambdaO1} and \eqref{dLambdaW1} in \eqref{dS1}.
The terms generated in $\d_{\L}S$ are not all linearly independent and their coefficients can not vanish
independently. This is resolved by the following identity of $\I_{\b\a(s)\ad(s)}$
\begin{IEEEeqnarray*}{ll}
    0=&~\frac{1}{(s+1)!}~\D_{(\a_{s+1}}\Dd^{\ad_{s}}\D_{\a_s}\I_{\a(s-1))\ad(s)}\n\label{Lambda1}\\
      &-\frac{s+1}{s}\frac{1}{(s+1)!}~\D_{(\a_{s+1}}\Dd^{\ad_s}\D^{\b}\I_{|\b|\a(s))\ad(s)}\\
      &+\frac{s+2}{s}\frac{1}{(s+1)!}~\D^2\Dd^{\ad_s}\I_{(\a_{s+1}\a(s))\ad(s)}
\end{IEEEeqnarray*}
which precisely recombines the above terms in linearly independent groups.
This is a higher spin generalization of a
corresponding identity found in \cite{Buchbinder:2021igw}. Using \eqref{Lambda1}, we can enforce
$\L$-invariance, $(\d_{\L}S=0)$ to find:
\begin{IEEEeqnarray*}{l}
    f_1=0~,~f_3-\frac{s+1}{s}~g_2+(\frac{s+1}{s})^2~d_1=0~,\n\\
    2f_4-\frac{s+2}{s}~g_2-\frac{s+1}{s}~g_3+2\frac{(s+1)(s+2)}{s^2}~d_1=0
\end{IEEEeqnarray*}
The system of equations \eqref{eta1} and \eqref{Lambda1} uniquely determines all coefficients up to an overall
scaling factor
\begin{IEEEeqnarray*}{ll}
    f_1=0~,~f_2=-c~,~f_3=2~\frac{s+1}{(s+2)^2}~c~,~f_4=c\n\\
    g_1=0~,~g_2=-2~\Big(\frac{s}{s+2}\Big)^2c~,~g_3=0~,\\
    d_1=-2~\Big(\frac{s}{s+2}\Big)^2c~,~d_2=0
\end{IEEEeqnarray*}
and fix the $3/2$-order invariant quantity $\J_{\b\a(s)\ad(s)}$:
\begin{IEEEeqnarray*}{l}
    \hspace{-8mm}\J_{\b\a(s)\ad(s)}\hspace{-1mm}=\hspace{-1mm}-c\Big[\Dd^2\I_{\b\a(s)\ad(s)}
                       -\frac{2(s+1)}{(s+2)^2}~\D_{\b}\Dd^\bd\bar{\I}_{\a(s)\bd\ad(s)}\n\\[-2pt]
        \hfill-~\Dd^\bd\D_{\b}\bar{\I}_{\a(s)\bd\ad(s)}
                       +\frac{2s^2}{(s+2)^2s!}~\D_\b\Dd_{(\ad_s}\bar{\I}_{\a(s)\ad(s-1))}\\[-4pt]
    \hfill+\frac{2s^2}{(s+2)^2s!s!}~C_{\b(\a_s}~\D^{\g}\Dd_{(\ad_s}\bar{\I}_{|\g|\a(s-1))\ad(s-1))}\Big]
\end{IEEEeqnarray*}
The equations of motion for $\O_{\b\a(s)\ad(s)}$ and $\W_{\b\a(s)\ad(s)}$ generated by \eqref{S1}
are
\begin{IEEEeqnarray*}{ll}\n
    \O_{\b\a(s)\ad(s)}:~&~ \W_{\b\a(s)\ad(s)}=-\J_{\b\a(s)\ad(s)}~,\sn\\[1pt]
    \W_{\b\a(s)\ad(s)}:~&~ \O_{\b\a(s)\ad(s)}=-\I_{\b\a(s)\ad(s)}~.\sn
\end{IEEEeqnarray*}
and they are consistent with transformations \eqref{dW1} and \eqref{dO1}. After integrating them out of action
\eqref{S1}, we find:
    \begin{IEEEeqnarray*}{ll}
S[\I]=\hspace{-1mm}c\hspace{-0.9mm}\int\hspace{-1mm} d^8z &\Big\{~\I^{\b\a(s)\ad(s)}~\Dd^2\I_{\b\a(s)\ad(s)}
    +c.c.\n\label{Snm}\\
&+\frac{2s^2}{(s+2)^2}~\I^{\b\a(s)\ad(s)}\D_{\b}\Dd_{\ad_s}\bar{\I}_{\a(s)\ad(s-1)}\hspace{-0.6mm}+\hspace{-1mm}c.c.\\
  &-\frac{4(s+1)}{(s+2)^2}~\I^{\b\a(s)\ad(s)}\D_{\b}\Dd^{\bd}\bar{\I}_{\a(s)\bd\ad(s)}\\
  &-2~\I^{\b\a(s)\ad(s)}\Dd^{\bd}\D_{\b}\bar{\I}_{\a(s)\bd\ad(s)}\\
  &+\frac{4s^2}{(s+2)^2}~\I^{\a(s-1)\ad(s)}\D^{\a_s}\Dd_{\ad_s}\bar{\I}_{\a(s)\ad(s-1)}\Big\}
    \end{IEEEeqnarray*}
By substituting equation \eqref{I1} in the above, we recover the action for massless, non-minimal, arbitrary
half-integer superspin supermultiplet $S_{(\Ysf=s+1/2)}~[H_{\a(s)\ad(s)},~\chi_{\a(s)\ad(s-1)}]$ as written in
\cite{Gates:2013rka} with the identification $H_{\a(s)\ad(s)}=\H_{\a(s)\ad(s)}+\bar{\H}_{\a(s)\ad(s)}$.

The equations of motion for superfields $\H_{\a(s)\ad(s)}$ and $\bar{\chi}_{\a(s-1)\ad(s)}$ are respectively:
\begin{IEEEeqnarray*}{ll}
    \E^{(\H)}_{\a(s)\ad(s)}=\D^{\b}\Dd^2\I_{\b\a(s)\ad(s)}+\Dd^{\bd}\D^2\bar{\I}_{\a(s)\bd\ad(s)}~,\n\\[2pt]
    \bar{\E}^{(\chi)}_{\a(s-1)\ad(s)}=\Dd^2\I_{\a(s-1)\ad(s)}
    +\Dd^{\bd}\D^{\a_s}\bar{\I}_{\a(s)\bd\ad(s)}\n\\[2pt]
   \hspace{20mm}+\frac{s}{(s+2)s!}~\D^{\a_s}\Dd_{(\ad_s}\bar{\I}_{\a(s)\ad(s-1))}\\[2pt]
    \hspace{20mm}+\frac{s+1}{s+2}~\D^{\a_s}\Dd^{\bd}\bar{\I}_{\a(s)\bd\ad(s)}
\end{IEEEeqnarray*}
and they satisfy the following Bianchi identities which capture the invariance under $L$, $\eta$ and $\L$
symmetries
\begin{IEEEeqnarray*}{ll}\n
\D^{\a_s}\E^{(\H)}_{\a(s)\ad(s)}=\D^2\bar{\E}^{(\chi)}_{\a(s-1)\ad(s)}~&~[L-\text{invariance}]\sn\label{LBI1}
\\[2pt]
    \E^{(\H)}_{\a(s)\ad(s)}=\bar{\E}^{(\H)}_{\a(s)\ad(s)}~&~[\eta-\text{invariance}]\sn\label{etaBI1}\\[2pt]
    \Dd_{(\ad_{s+1}}\bar{\E}^{(\chi)}_{\a(s-1)\ad(s))}=0~&~[\L-\text{invariance}]~~~\sn\label{LambaBI1}
\end{IEEEeqnarray*}
Identities \eqref{LBI1} and \eqref{LambaBI1} are not new as they have the same form as the Bianchi identities
of the $S[H_{\a(s)\ad(s)},\chi_{\a(s)\ad(s-1)}]$ theory. However, they acquire a new interpretation due to the
existence of this half-order formulation of the theory.
Equation \eqref{LBI1} is a consequence of the ability to write the
action purely in terms of $\I_{\b\a(s)\ad(s)}$.
Identity \eqref{LambaBI1} is just a repackaging of property \eqref{Lambda1} of this basic building block.
Identity \eqref{etaBI1} ---which was previously trivially true
because superfield $H_{\a(s)\ad(s)}$ is real--- is now understood as the Bianchi identity for a
new local symmetry \eqref{detaH} which provides a higher spin generalization off linearized changed of
coordinates.
\subsection{First order formalism of minimal half-integer superspin supermultiplets}
For the $A$=$-1$ case and according to equations \eqref{mI} and \eqref{mK} there are two
invariants. Equation \eqref{mI} gives the first-order invariant $\I_{\b\a(s)\ad(s-1)}$ and \eqref{mK} gives the
half-order invariant $K_{\a(s+1)\ad(s)}$. Based on their respective definitions, it is clear that the
symmetric part of $\I_{\b\a(s)\ad(s-1)}$ does not capture new information since it can be written as the
derivative of $\K_{\a(s+1)\ad(s)}$:
\begin{equation}
    \frac{1}{(s+1)!}~\I_{(\b\a(s))\ad(s-1)}=\Dd^{\ad_s}\K_{\b\a(s)\ad(s}
\end{equation}
However, the antisymmetric part $\I_{\a(s-1)\ad(s-1}$ defined as:
\begin{IEEEeqnarray*}{l}
    \I_{\a(s-1)\ad(s-1}= C^{\b\a_s}\I_{\b\a(s)\ad(s-1)}\n\label{I2}\\
            \hspace{15mm}=-\Dd^{\ad_s}\D^{\a_s}\H_{\a(s)\ad(s)}-\frac{s+1}{s}\D^{\a_s}\Dd^{\ad_s}\H_{\a(s)\ad(s)}\\
                     \hspace{17mm}~-\frac{s+1}{s}\frac{1}{(s-1)!}\D_{(\a_{s-1}}\bar{\chi}_{\a(s-2))\ad(s-1)}
\end{IEEEeqnarray*}
is an independent first-order invariant. Therefore, we consider the $S$=$S[\I,\K]$ class of actions
which are functionals of $\I_{\a(s-1)\ad(s-1)}$ and $\K_{\a(s+1)\ad(s)}$ and thus automatically $L$-invariant.

By construction $\I_{\a(s-1)\ad(s-1)}$ and $\K_{\a(s+1)\ad(s)}$ are also invariant under the
$\L$-redundancy of $\chi_{\a(s-1)\ad(s-2)}$:
\begin{IEEEeqnarray*}{ll}\n
\d_{\L}\chi_{\a(s-1)\ad(s-2)}=\frac{1}{(s-2)!}$$\Dd_{(\ad_{s-2}}\L_{\a(s-1)\ad(s-3))}~,~~~~~\sn\\[1pt]
\d_{\L}\I_{\b\a(s)\ad(s-1)}=0\sn~,\\[1pt]
\d_{\L}\K_{\a(s+1)\ad(s)}=0\sn~.
\end{IEEEeqnarray*}
However, under $\eta$-symmetry \eqref{cH} they transform as:
\begin{IEEEeqnarray*}{ll}\n
    \d_{\eta}\I_{\a(s-1)\ad(s-1)}=&-i~\Dd^{\ad_s}\D^{\a_s}\eta_{\a(s)\ad(s)}\sn\\
                                  &-i~\frac{s+1}{s}~\D^{\a_s}\Dd^{\ad_s}\eta_{\a(s)\ad(s)}~,\\
    \d_{\eta}\K_{\a(s+1)\ad(s)}=&~\frac{i}{(s+1)!}~\D_{(\a_{s+1}}\eta_{\a(s))\ad(s)}~.\sn
\end{IEEEeqnarray*}

These symmetries will be imposed on the action via three auxiliary superfields. One auxiliary superfield
$\omega_{\a(s-1)\ad(s-1)}$  will couple to the first order invariant $\I_{\a(s-1)\ad(s-1)}$ and two more
auxiliary superfields ($\W_{\a(s+1)\ad(s)},~\O_{\a(s+1)\ad(s)}$) for the half-order invariant
$\K_{\a(s+1)\ad(s)}$. The general action takes the form:
\begin{IEEEeqnarray*}{ll}
    S=\int d^8z\Big\{
    &c_1~\omega^{\a(s-1)\ad(s-1)}\omega_{\a(s-1)\ad(s-1)}\n\label{S2}\\
&+\frac{c_2}{2}~\omega^{\a(s-1)\ad(s-1)}\bar{\omega}_{\a(s-1)\ad(s-1)}\\
&+\omega^{\a(s-1)\ad(s-1)}\I_{\a(s-1)\ad(s-1)}\\
&+\W^{\a(s+1)\ad(s)}\Omega_{\a(s+1)\ad(s)}\\
&+\W^{\a(s+1)\ad(s)}\K_{\a(s+1)\ad(s)}\\
&+\O^{\a(s+1)\ad(s)}\J_{\a(s+1)\ad(s)}\Big\}+c.c.
\end{IEEEeqnarray*}
where $\J_{\a(s+1)\ad(s)}$  is a $3/2$-order invariant constructed out of the derivatives of
$\K_{\a(s+1)\ad(s)}$:
\begin{IEEEeqnarray*}{ll}
    \J_{\a(s+1)\ad(s)}=&~d_1~\K_{\a(s+1)\ad(s)}\n\\
                       &+\frac{d_2}{(s+1)!}~\Dd^{\ad_{s+1}}\D_{(\a_{s+1}}\bar{\K}_{\a(s))\ad(s+1)}\\
                       &+\frac{d_3}{(s+1)!}~\D_{(\a_{s+1}}\Dd^{\ad_{s+1}}\bar{\K}_{\a(s))\ad(s+1)}
\end{IEEEeqnarray*}
The transformations of the three auxiliary superfields are chosen such that the variation of action \eqref{S2}
is simplified by eliminating all the terms proportional to $\omega_{\a(s-1)\ad(s-1)}, \O_{\a(s+1)\ad(s)}$ and
$\W_{\a(s+1)\ad(s)}$ ---similar to \eqref{dS1}:
\begin{IEEEeqnarray*}{ll}
    \d\omega_{\a(s-1)\ad(s-1)}=&-\frac{2c_1}{4c_1{}^2-c_2{}^2}~\d\I_{\a(s-1)\ad(s-1)}\n\\
                               &+\frac{c_2}{4c_1{}^2-c_2{}^2}~\d\bar{\I}_{\a(s-1)\ad(s-1)}~,\\[3pt]
    \d_{L}\omega_{\a(s-1)\ad(s-1)}=&0~,\sn\\[1pt]
    \d_{\L}\omega_{\a(s-1)\ad(s-1)}=&0~,\sn\\[1pt]
\d_{\eta}\omega_{\a(s-1)\ad(s-1)}=&~i~\frac{2c_1-\frac{s+1}{s}c_2}{4c_1{}^2-c_2{}^2}~
\Dd^{\ad_{s}}\D^{\a_s}\eta_{\a(s)\ad(s)}~~~~~\sn\label{detao}\\[2pt]
&+i~\frac{2\frac{s+1}{s}c_1-c_2}{4c_1{}^2-c_2{}^2}~
\D^{\a_{s}}\Dd^{\ad_s}\eta_{\a(s)\ad(s)}
\end{IEEEeqnarray*}
\begin{IEEEeqnarray*}{ll}
    \d\O_{\a(s+1)\ad(s)}=&-\d\K_{\a(s+1)\ad(s)}~,\n\\[3pt]
    \d_{L}\O_{\a(s+1)\ad(s)}=&0~,\sn\\[1pt]
    \d_{\L}\O_{\a(s+1)\ad(s)}=&0~,\sn\\[1pt]
\d_{\eta}\O_{\a(s+1)\ad(s)}=&-\frac{i}{(s+1)!}~\D_{(\a_{s+1}}\eta_{\a(s))\ad(s)}\sn\label{detaO2}
\end{IEEEeqnarray*}
\begin{IEEEeqnarray*}{ll}
    \d\W_{\a(s+1)\ad(s)}=&-\d\J_{\a(s+1)\ad(s)}~,\n\\[3pt]
    \d_{L}\W_{\a(s+1)\ad(s)}=&0~,\sn\\[1pt]
    \d_{\L}\W_{\a(s+1)\ad(s)}=&0~,\sn\\[1pt]
\d_{\eta}\W_{\a(s+1)\ad(s)}=&-i~\frac{d_1+d_2}{(s+1)!}~\Dd^2\D_{(\a_{s+1}}\eta_{\a(s))\ad(s)}\sn\label{detaW2}\\[2pt]
&-i~\frac{d_2-\frac{s+2}{s+1}d_3}{(s+1)!}~\D_{(\a_{s+1}}\Dd^2\eta_{\a(s))\ad(s)}\\[2pt]
&+i~\frac{\frac{s}{s+1}d_2}{(s+1)!}~\Dd_{(\ad_s}\D_{(\a_{s+1}}\Dd^{\gd}\eta_{\a(s))|\gd|\ad(s-1))}
\end{IEEEeqnarray*}
Using equations \eqref{detao}, \eqref{detaO2} and \eqref{detaW2}, we find that $\eta$-invariance of action
\eqref{S2} requires:
\begin{IEEEeqnarray*}{l}
c_1=c~,~c_2=\frac{2s}{s+1}~c~,~d_1=\frac{1}{4c}~\Big(\frac{s+1}{s}\Big)^3~,\n\\[2pt]
d_2=\frac{1}{4c}~\Big(\frac{s+1}{s}\Big)^3~,~d_3=\frac{1}{2c}~\frac{(s+1)^4}{s^3(s+2)^2}
\end{IEEEeqnarray*}
which fix the $3/2$-invariant $\J_{\a(s+1)\ad(s)}$ to be:
\begin{IEEEeqnarray*}{l}
    \J_{\a(s+1)\ad(s)}=-\frac{1}{4c}~\Big(\frac{s+1}{s}\Big)^3~\Dd^2\K_{\a(s+1)\ad(s)}\n\\[2pt]
\hfill+\frac{1}{4c(s+1)!}\Big(\frac{s+1}{s}\Big)^3\Dd^{\ad_{s+1}}\D_{(\a_{s+1}}\bar{\K}_{\a(s))\ad(s+1)}\\[2pt]
\hfill+\frac{1}{2c(s+1)!}~\frac{(s+1)^4}{s^3(s+2)^2}\D_{(\a_{s+1}}\Dd^{\ad_{s+1}}\bar{\K}_{\a(s))\ad(s+1)}
\end{IEEEeqnarray*}
The equations of motion for the auxiliary superfields are:
\begin{IEEEeqnarray*}{ll}
    \omega_{\a(s-1)\ad(s-1)}=&-\frac{1}{2c}~\frac{(s+1)^2}{2s+1}~\I_{\a(s-1)\ad(s-1)}\n\\[2pt]
                             &+\frac{1}{2c}~\frac{s(s+1)}{2s+1}~\bar{\I}_{\a(s-1)\ad(s-1)}~,\\[3pt]
    \O_{\a(s+1)\ad(s)}=&-\K_{\a(s+1)\ad(s)}~,\n\\[3pt]
    \W_{\a(s+1)\ad(s)}=&-\J_{\a(s+1)\ad(s)}~.\n
    \end{IEEEeqnarray*}
Because of their algebraic nature, we can integrate them out of \eqref{S2} in order to find the $S[\I,\K]$
action:
\begin{IEEEeqnarray*}{l}
    S=\hspace{-1mm}\frac{1}{4c}\hspace{-1mm}\int\hspace{-1mm} d^8z\Bigg\{
        \Big(\frac{s+1}{s}\Big)^3\K^{\a(s+1)\ad(s)}\Dd^2\K_{\a(s+1)\ad(s)}\hspace{-1mm}+\hspace{-1mm}c.c.~~~~~\n\label{mS}\\
     \hfill-2~\Big(\frac{s+1}{s}\Big)^3\K^{\a(s+1)\ad(s)}\Dd^{\ad_{s+1}}\D_{\a_{s+1)}}\bar{\K}_{\a(s)\ad(s+1)}\\
     \hfill-4~\frac{(s+1)^4}{s^3(s+2)^2}~\K^{\a(s+1)\ad(s)}\D_{\a_{s+1)}}\Dd^{\ad_{s+1}}\bar{\K}_{\a(s)\ad(s+1)}\\
     \hfill-~\frac{(s+1)^2}{2s+1}~\I^{\a(s-1)\ad(s-1)}\I_{\a(s-1)\ad(s-1)}+c.c.\\
     \hfill+2~\frac{s(s+1)}{2s+1}~\I^{\a(s-1)\ad(s-1)}\bar{\I}_{\a(s-1)\ad(s-1)}\Bigg\}
\end{IEEEeqnarray*}
By substituting equations \eqref{mK} and \eqref{I2} back in the above action, we recover precisely the minimal
half-integer superspin supermultiplet action $S_{(\Ysf=s+1/2)}[H_{\a(s)\ad(s), \chi_{\a(s-1)\ad(s-2)}}]$ as
written in \cite{Gates:2013rka} with the identification
$H_{\a(s)\ad(s)}=\H_{\a(s)\ad(s)}+\bar{\H}_{\a(s)\ad(s)}$.

The equations of motion for superfields $\H_{\a(s)\ad(s)}$ and $\bar{\chi}_{\a(s-2)\ad(s-1)}$
are respectively:
\begin{IEEEeqnarray*}{l}
    \E^{(\H)}_{\a(s)\ad(s)}\hspace{-0.8mm}=\frac{2s+1}{s^2}\hspace{-0.8mm}\Big[\D^{\b}\Dd^2\K_{\b\a(s)\ad(s)}
        \hspace{-1mm}+\hspace{-1mm}\Dd^{\bd}\D^2\bar{\K}_{\a(s)\bd\ad(s)}\Big]\n\\[2pt]
\hfill+\frac{1}{s!s!}~\D_{(\a_s}\Dd_{(\ad_s}
\Big[\frac{s}{s+1}\I_{\a(s-1))\ad(s-1))}-\bar{\I}_{\a(s-1))\ad(s-1))}\Big]\\[2pt]
\hfill-\frac{1}{s!s!}~\Dd_{(\ad_s}\D_{(\a_s}
\Big[\frac{s}{s+1}\bar{\I}_{\a(s-1))\ad(s-1))}-\I_{\a(s-1))\ad(s-1))}\Big]\\[10pt]
\bar{\E}^{(\chi)}_{\a(s-2)\ad(s-1)}=\D^{\a_{s-1}}\Big[
\frac{s}{s+1}~\bar{\I}_{\a(s-1)\ad(s-1)}\n\\
\hspace{42mm}-~\I_{\a(s-1)\ad(s-1)}\Big]
\end{IEEEeqnarray*}
and they satisfy the following Bianchi identities for the invariance under $L$, $\eta$ and $\L$
symmetries:
\begin{IEEEeqnarray*}{l}\n
0=\D^{\a_s}\E^{(\H)}_{\a(s)\ad(s)}\hfill[L-\text{invariance}]~~~~\sn\label{LBI2}\\
\hspace{6mm}+\frac{1}{s!(s-1)!}~\Dd_{(\ad_s}\D_{(\a_{s-1}}\bar{\E}^{(\chi)}_{\a(s-2))\ad(s-1))}\\
\hspace{6mm}+\frac{s-1}{s}\frac{1}{s!(s-1)!}~\D_{(\a_{s-1}}\Dd_{(\ad_s}\bar{\E}^{(\chi)}_{\a(s-2))\ad(s-1))}~,\\[4pt]
    \E^{(\H)}_{\a(s)\ad(s)}=\bar{\E}^{(\H)}_{\a(s)\ad(s)}~,\hfill[\eta-\text{invariance}]~~~~\sn\label{etaBI2}\\[4pt]
    \D^{\a_{s-2}}\bar{\E}^{(\chi)}_{\a(s-2)\ad(s-1)}=0~.\hfill[\L-\text{invariance}]~~~~\sn\label{LambaBI2}
\end{IEEEeqnarray*}
Identities \eqref{LBI2} and \eqref{LambaBI2} are consequences
of the action being written purely in terms of the $L$ and $\L$-invariant half and first order
quantities $\K_{\a(s+1)\ad(s)}$ and $\I_{\a(s-1)\ad(s-1)}$.
\section{Integer Superspin}
The integer superspin supermultiplet $\Ysf=s$ on-shell describes the propagation of the massless spins
$j=s+1/2$ and $j=s$. Its superspace realization is given by the
equivalence class $[\Psi_{\a(s)\ad(s-1)}]$ of a $(s,s-1)$ SL(2,$\mathbb{C}$)  superfield  tensor
$\Psi_{\a(s)\ad(s-1)}$ which is independently symmetric in both types of spinorial indices.
The superfield strength that describes the physical degrees of freedom, constructed in \cite{Gates:1983nr},
is:
\begin{IEEEeqnarray*}{l}
W_{\a(2s)}\hspace{-0.8mm}=
\Dd^2\D_{(\a_{2s}}\pa_{\a_{2s-1}}{}^{\ad_{1}}\ldots\pa_{\a_{s+1}}{}^{\ad_{s-1}}\Psi_{\a(s))\ad(s-1)}~~~~\n
\end{IEEEeqnarray*}
For $s>1$, the general redundancy which respects this superfield strength and defines the equivalence class
is\footnote{For $s=1$ we have $\hat{\Psi}_\a\sim\Psi_{\a}+\D_{\a}l+\Dd^2\L_{\a}$}:
\begin{IEEEeqnarray*}{ll}
\hat{\Psi}_{\a(s)\ad(s-1)}\sim\Psi_{\a(s)\ad(s-1)}&+\tfrac{1}{s!}~\D_{(\a_s}l_{\a(s-1))\ad(s-1)}\n\label{dPsi}\\[1pt]
&+\tfrac{1}{(s-1)!}~\Dd_{(\ad_{s-1}}\L_{\a(s)\ad(s-2))}
\end{IEEEeqnarray*}

In this case, the highest propagating spin is a fermion and thus its equation of motion is the,
higher spin generalization of the massless Dirac equation. In general, for these equations of motion
the Ostrogradskiy procedure can not be applied since they are already first order and the partial derivative
operator can not be further factorized. However, as mentioned previously, for supersymmetric theories such
factorization of first order differential operators is possible due to the supersymmetry algebra. The Dirac
equation for $\Psi_{\a(s)\ad(s-1)}$ is $\pa_{(\a_{s}}{}^{\ad_s}\bar{\Psi}_{\a(s-1))\ad(s)}$=$0$. In superspace,
this will include terms $\Dd^{\ad_{s}}\D_{(\a_{s}}\bar{\Psi}_{\a(s-1))\ad(s)}$ and
$\D_{(\a_{s}}\Dd^{\ad_{s}}\bar{\Psi}_{\a(s-1))\ad(s)}$.
Therefore, we consider the following two candidates for half-order invariants:
\begin{IEEEeqnarray*}{l}\n
    I_{\a(s)\bd\ad(s-1)}=\Dd_{(\bd}\Psi_{\a(s)\ad(s-1))}~,\sn\label{psiI1}\\
    I_{\a(s-1)\ad(s-1)}=\D^{\b}\Psi_{\b\a(s-1)\ad(s-1)}~.\sn\label{psiI2}
\end{IEEEeqnarray*}

For $s=1$, the real part of \eqref{psiI2} $\big\{2Re[I]=\D^{\a}\Psi_{\a}+\Dd^{\ad}\bar{\Psi}_{\ad}\big\}$
and also
$I_{\a\bd}$ of \eqref{psiI1} are $\L$-invariant for an appropriate reduction of the $\L_{\a}$ symmetry
$\L_{\a}=i\D_{\a}\L~,~\L=\bar{\L}$. Following this will lead to the
half-order formulation of the Ogievetsky-Sokatchev description of the $(3/2,1)$ supermultiplet
\cite{Ogievetsky:1975vk}. However, this can not be generalized to $s>1$. An alternative approach that can be
extended to higher spin supermultiplets is to instead restrict the $l_{\a(s-1)\ad(s-1)}$ symmetry:
\begin{equation}
l_{\a(s-1)\ad(s-1)}=\D^{\a_{s}}L_{\a(s)\ad(s-1)}~.
\end{equation}
By introducing a complex, bosonic, compensating superfield $\V_{\a(s-1)\ad(s-1)}$
with transformation $\d_{L}\V_{\a(s-1)\ad(s-1)}=\D^{\a_{s}}L_{\a(s)\ad(s-1)}$, we find an elementary
half-order, $L$-invariant building block:
\begin{IEEEeqnarray*}{l}
    \I_{\b\a(s-1)\ad(s-1)}=\Psi_{\b\a(s-1)\ad(s-1)}-\D_{\b}\V_{\a(s-1)\ad(s-1)}.~~~~~\n\label{I3}
\end{IEEEeqnarray*}
Under the $\L$-redundancy, $\I_{\b\a(s-1)\ad(s-1)}$ transforms as:
\begin{IEEEeqnarray*}{l}
\d_{\L}\I_{\b\a(s-1)\ad(s-1)}=\tfrac{1}{(s-1)!}~\Dd_{(\ad_{s-1}}\L_{\b\a(s-1)\ad(s-2))}~.~~~~
\n\label{dLambdaI3}
\end{IEEEeqnarray*}

Moreover compensator $\V_{\a(s-1)\ad(s-1)}$, in addition to its $L$-transformation, it can be assigned
a local algebraic symmetry similar in nature to \eqref{detaH}
\begin{IEEEeqnarray*}{l}\n
\d_{\eta}\V_{\a(s-1)\ad(s-1)}=i~\eta_{\a(s-1)\ad(s-1)}~,\sn\label{detaV}\\
\d_{\eta}\I_{\b\a(s-1)\ad(s-1)}=-i~\D_{\b}\eta_{\a(s-1)\ad(s-1)}\sn\label{detaI3}
\end{IEEEeqnarray*}
with a real parameter $\eta_{\a(s-1)\ad(s-1)}$=$\bar{\eta}_{\a(s-1)\ad(s-1)}$.  Similar to the discussion in
the previous section, the class of actions $S$=$S[\I]$ is trivially $L$-invariant and the invariance with
respect to symmetries \eqref{dLambdaI3} and \eqref{detaI3} will be enforced by two auxiliary
superfields $\W_{\b\a(s-1)\ad(s-1)}$, $\O_{\b\a(s-1)\ad(s-1)}$. The general action has the form
\begin{IEEEeqnarray*}{ll}
    S=\int d^8z\Big\{&\W^{\b\a(s-1)\ad(s-1)}\O_{\b\a(s-1)\ad(s-1)}\n\label{S3}\\[2pt]
        &+\W^{\b\a(s-1)\ad(s-1)}\I_{\b\a(s-1)\ad(s-1)}\\[2pt]
        &+\O^{\b\a(s-1)\ad(s-1)}\J_{\b\a(s-1)\ad(s-1)}\Big\}+c.c.
\end{IEEEeqnarray*}
where $\J_{\b\a(s-1)\ad(s-1)}$ is a general $3/2$-order invariant constructed out
of the derivatives of $\I_{\b\a(s-1)\ad(s-1)}$:
\begin{widetext}
\begin{IEEEeqnarray*}{l}
    \J_{\b\a(s-1)\ad(s-1)}=c_1~\D^2\I_{\b\a(s-1)\ad(s-1)}+c_2~\Dd^2\I_{\b\a(s-1)\ad(s-1)}
    +c_3~\D_{\b}\Dd^{\bd}\bar{\I}_{\a(s-1)\bd\ad(s-1)}+c_4~\Dd^{\bd}\D_{\b}\bar{\I}_{\a(s-1)\bd\ad(s-1)}~~~\n\\
   \hfill+\tfrac{d_1}{(s-1)!}~C_{\b(\a_{s-1}}\Dd^2\I_{\a(s-2))\ad(s-1)}
   +\tfrac{d_2}{(s-1)!}~\D_{\b}\Dd_{(\ad_{s-1}}\bar{\I}_{\a(s-1)\ad(s-2))}
   +\tfrac{d_3}{(s-1)!}~\Dd_{(\ad_{s-1}}\D_{\b}\bar{\I}_{\a(s-1)\ad(s-2))}\\
   \hfill+\tfrac{f_1}{(s-1)!(s-1)!}~C_{\b(\a_{s-1}}\D^{\g}\Dd_{(\ad_{s-1}}\bar{\I}_{|\g|\a(s-2))\ad(s-2)}
   +\tfrac{f_2}{(s-1)!(s-1)!}~C_{\b(\a_{s-1}}\Dd_{(\ad_{s-1}}\D^{\g}\bar{\I}_{|\g|\a(s-2))\ad(s-2)}
\end{IEEEeqnarray*}
\end{widetext}
and
\begin{IEEEeqnarray*}{ll}
    \I_{\a(s-2)\ad(s-1)}&=C^{\b\a_{s-1}}\I_{\b\a(s-1)\ad(s-1)}\n\\
    &=\D^{\a_{s-1}}\V_{\a(s-1)\ad(s-1}
\end{IEEEeqnarray*}

Action \eqref{S3} is identical in form with action \eqref{S1}, hence its variation will take the same form as
\eqref{dS1}. The transformation laws for $\W_{\b\a(s-1)\ad(s-1)}$ and $\O_{\b\a(s-1)\ad(s-1)}$ are then
chosen in a similar manner, by eliminating the terms of the variation which are proportional to the
auxiliary superfields. Therefore for $\O_{\b\a(s-1)\ad(s-1)}$ we get:
\begin{IEEEeqnarray*}{ll}
    \d\O_{\b\a(s-1)\ad(s-1)}=-\d\I_{\b\a(s-1)\ad(s-1)}~,\n\\[3pt]
    \d_{L}\O_{\b\a(s-1)\ad(s-1)}=0~,\sn\\[2pt]
\d_{\L}\O_{\b\a(s-1)\ad(s-1)}=-\tfrac{1}{(s-1)!}~\Dd_{(\ad_{s-1}}\L_{\b\a(s-1)\ad(s-2))}~,\sn\label{dLambdaO3}\\[2pt]
    \d_{\eta}\O_{\b\a(s-1)\ad(s-1)}=i~\D_{\b}\eta_{\a(s-1)\ad(s-1)}~,\sn\label{detaO3}
\end{IEEEeqnarray*}
and likewise for $\W_{\b\a(s-1)\ad(s-1)}$:
\begin{widetext}
\vspace*{-4mm}
\begin{IEEEeqnarray*}{ll}
    \d\W_{\b\a(s-1)\ad(s-1)}=&-\d\J_{\b\a(s-1)\ad(s-1)}~,\n\\[3pt]
    \d_{L}\W_{\b\a(s-1)\ad(s-1)}=&0~,\sn\\[1pt]
    \d_{\L}\W_{\b\a(s-1)\ad(s-1)}=&-\frac{c_1}{(s-1)!}~\D^2\Dd_{(\ad_{s-1}}\L_{\b\a(s-1)\ad(s-2))}
    -\frac{c_3}{(s-1)!}~\D_\b\Dd^{\bd}\D_{(\a_{s-1}}\bar{\L}_{\a(s-2))\bd\ad(s-1)}\sn\label{dLambdaW3}\\
    &+\frac{c_4}{(s-1)!}~C_{\b(\a_{s-1}}~\Dd^{\bd}\D^2\bar{\L}_{\a(s-2))\bd\ad(s-1)}~,\\[1pt]
    \d_{\eta}\W_{\b\a(s-1)\ad(s-1)}=&i(c_2+c_4)~\Dd^2\D_{\b}\eta_{\a(s-1)\ad(s-1)}
    +i(-2c_3+c_4+d_2)~\D_{\b}\Dd^2\eta_{\a(s-1)\ad(s-1)}\sn\label{detaW3}\\
    &-i\frac{d_3}{(s-1)!}~\Dd_{(\ad_{s-1}}\D_{\b}\Dd^{\gd}\eta_{\a(s-1)|\gd|\ad(s-2))}
-i\frac{d_1}{(s-1)!}~C_{\b(\a_{s-1}}~\Dd^2\D^{\g}\eta_{|\g|\a(s-2))\ad(s-1)}\\
    &+i\frac{f_1}{(s-1)!}~C_{\b(\a_{s-1}}~\D^{\g}\Dd^2\eta_{|\g|\a(s-2))\ad(s-1)}
    -i\frac{f_2}{(s-1)!}~C_{\b(\a_{s-1}}~\Dd_{(\ad_{s-1}}\D^{\g}\Dd^{\gd}\eta_{|\g|\a(s-2))|\gd|\ad(s-2))}
\end{IEEEeqnarray*}
\end{widetext}

Using equations \eqref{detaW3} and \eqref{detaO3} we find that the $\eta$-invariance of action \eqref{S3} requires
\begin{IEEEeqnarray*}{l}
    c_2+c_4=0~,~d_2=4c_3-2c_4~,~d_1=0~,\n\label{Sol31}\\
    d_3=0~,~f_1=d_2~,~f_2=0
\end{IEEEeqnarray*}
For $\L$-invariance, some of the terms generated in the variation of the action by substituting
\eqref{dLambdaW3} and \eqref{dLambdaO3} vanish due to the following identities:
\begin{equation}\label{id3}
    \D^{\a_{s-2}}\I_{\a(s-2)\ad(s-1)}=0~,~\D^2\I_{\a(s-2)\ad(s-1)}=0~.
\end{equation}
The rest are eliminated by choosing the coefficients:
\begin{equation}\label{Sol32}
c_1=0~,~c_3=0~,~d_3=\frac{s-1}{s}(d_2+2c_4)
\end{equation}
The two systems \eqref{Sol31} and \eqref{Sol32} give
\begin{IEEEeqnarray*}{l}
    c_1=0~,~c_2=-c~,~c_3=0~,~c_4=c~,\n\\
    d_1=0~,~d_2=-2c~,~d_3=0~,~f_1=-2c~,~f_2=0
\end{IEEEeqnarray*}
and determine $\J_{\b\a(s-1)\ad(s-1)}$
\begin{IEEEeqnarray*}{l}
    \J_{\b\a(s-1)\ad(s-1)}=\\
=-c~\Big[\Dd^2\I_{\b\a(s-1)\ad(s-1)}
-~\Dd^{\bd}\D_{\b}\bar{\I}_{\a(s-1)\bd\ad(s-1)}\n\\
\hspace{12mm}+\frac{2}{(s-1)!}~\D_{\b}\Dd_{(\ad_{s-1}}\bar{\I}_{\a(s-1)\ad(s-2))}\\
\hspace{12mm}+\frac{2}{[(s-1)!]^2}~C_{\b(\a_{s-1}}\D^{\g}\Dd_{(\ad_{s-1}}\bar{\I}_{|\g|\a(s-2))\ad(s-2)}\Big]
\end{IEEEeqnarray*}
We integrate out the auxiliary superfields $\W_{\b\a(s-1)\ad(s-1)}$ and $\O_{\b\a(s-1)\ad(s-1)}$ using their
respective equations of motion
\begin{IEEEeqnarray*}{ll}\n
    \W_{\b\a(s-1)\ad(s-1)}=-\J_{\b\a(s-1)\ad(s-1)}~,\sn\\[1pt]
    \O_{\b\a(s-1)\ad(s-1)}=-\I_{\b\a(s-1)\ad(s-1)}\sn
\end{IEEEeqnarray*}
to find
the action $S[\I]$ written purely in terms of $\I_{\b\a(s-1)\ad(s-1)}$:
\begin{IEEEeqnarray*}{ll}
    S=c\hspace{-1mm}\int d^8z\Big\{&\I^{\b\a(s-1)\ad(s-1)}\Dd^2\I_{\b\a(s-1)\ad(s-1)}+c.c.\n\label{s3}\\
                      &-2~\I^{\b\a(s-1)\ad(s-1)}\Dd^{\bd}\D_{\b}\bar{\I}_{\a(s-1)\bd\ad(s-1)}\\
  &+2~\I^{\b\a(s-1)\ad(s-1)}\D_{\b}\Dd_{\ad_{s-1}}\bar{\I}_{\a(s-1)\ad(s-2)}\hspace{-1mm}+\hspace{-1mm}c.c.\\
          &+4~\I^{\a(s-2)\ad(s-1)}\D^{\a_{s-1}}\Dd_{\ad_{s-1}}\bar{\I}_{\a(s-1)\ad(s-2)}\Big\}
\end{IEEEeqnarray*}
By substituting \eqref{I3}, this action becomes exactly the integer superspin action
$S_{(\Ysf=s)}[\Psi_{\a(s)\ad(s-1)},V_{\a(s-1)\ad(s-1)}]$ of \cite{Gates:2013rka} with the identification
$V_{\a(s-1)\ad(s-1)}$ = $\V_{\a(s-1)\ad(s-1)}+\bar{\V}_{\a(s-1)\ad(s-1)}$.

The equations of motion for superfields $\Psi_{\a(s)\ad(s-1)}$ and $\V_{\a(s-1)\ad(s-1)}$ are respectively:
\begin{IEEEeqnarray*}{ll}
    \E^{(\Psi)}_{\a(s)\ad(s-1)}=&\frac{1}{s!}~\Dd^2\I_{(\a_s\a(s-1))\ad(s-1)}\n\\
                                &-\frac{1}{s!}~\Dd^{\bd}\D_{(\a_s}\bar{\I}_{\a(s-1))\bd\ad(s-1)}\\
                                &+\frac{1}{s!(s-1)!}~\D_{(\a_s}\Dd_{(\ad_{s-1}}\bar{\I}_{\a(s-1))\ad(s-2))}
\end{IEEEeqnarray*}
and
\begin{IEEEeqnarray*}{ll}
    \E^{(\V)}_{\a(s-1)\ad(s-1)}=&~~\D^{\b}\Dd^2\I_{\b\a(s-1)\ad(s-1)}\n\\
                                             &+\Dd^{\bd}\D^2\bar{\I}_{\a(s-1)\bd\ad(s-1)}
\end{IEEEeqnarray*}
It is straightforward to show that they satisfy the following identities which play the role of
Bianchi identities for the invariance under $L$, $\eta$ and $\L$ symmetries
\begin{IEEEeqnarray*}{ll}\n
\D^2\E^{(\Psi)}_{\a(s)\ad(s-1)}+\frac{1}{s!}~\D_{(\a_s}\E^{(\V)}_{\a(s-1))\ad(s-1)}=0~,&\sn\label{LBI3}
\\[2pt]
    \E^{(\V)}_{\a(s-1)\ad(s-1)}=\bar{\E}^{(\V)}_{\a(s-1)\ad(s-1)}~,&\sn\label{etaBI3}\\[2pt]
    \Dd_{(\ad_{s-1}}\E^{(\Psi)}_{\a(s)\ad(s-1)}=0~.&~~~~~~~\sn\label{LambaBI3}
\end{IEEEeqnarray*}

The interpretation of this equations is now clear. Identity \eqref{LBI3} reflects that the half-order
description of this system exist and the action for integer superspins can be written purely in terms of a
basic building block
$\I_{\b\a(s-1)\ad(s-1)}$. Equation \eqref{LambaBI3} is a repackaging of the properties \eqref{id3} of
$\I_{\b\a(s-1)\ad(s-1)}$. Finally, equation \eqref{etaBI3} is a manifestation of the additional local symmetry
\eqref{detaV}.
\section{Summary \& Conclusions}
In this work, we have shown that higher spin supermultiplets can be described using a first or even half order
formalism. Following the conventional first-order approach, we start with the set of unconstrained
superfields that participate in the Lagrangian description of the free theories and we relax some of their
characteristic properties. For non-supersymmetric higher spin fields one relaxes the symmetry of their
indices.  For higher spin superfields this corresponds to relaxing their reality and corresponding
gauge redundancies.

In this configuration, we find first and half order gauge invariants which are used to write trivially
invariant actions. However, in order to maintain the degrees of freedom of the theory a new local symmetry is
introduced. This symmetry is a higher spin generalization of the linearized general covariance in supergravity
which reduces the complex prepotential to a real one. Checking the invariance of the action under this new
symmetry is more involved but it is possible due to various identities of the basic invariant building blocks.
These identities are promoted to play the role of Bianchi identities.

The procedure is simplified by introducing auxiliary superfields that play the role of connections in the
sense that they impose the new local symmetry and their equations of motion are algebraic in nature that
allows their integration out of the action.

For half-integer superspins ($s>0$) we find two varieties of first order description. In the first one, there
is a fundamental half-order building block $\I_{\b\a(s)\ad(s)}$ \eqref{I1} and the action \eqref{S1} requires
a pair of auxiliary superfields $\O_{\b\a(s)\ad(s)}$ and $\W_{\b\a(s)\ad(s)}$. Their integration yields action
\eqref{Snm} which is equivalent to the action for the non-minimal half-integer superspin supermultiplet.  This
result generalizes to higher spins the first order description of non-minimal supergravity supermultiplet
found in \cite{Buchbinder:2021igw}.  The second variety, allows for the construction of two independent
building blocks, $\I_{\a(s-1)\ad(s-1)}$ \eqref{I2} and $\K_{\a(s+1)\ad(s)}$ \eqref{mK} which are first and
half order respectively. The action \eqref{S2} requires three auxiliary superfields
$\omega_{\a(s-1)\ad(s-1)}$, $\O_{\a(s+1)\ad(s)}$ and $\W_{\a(s+1)\ad(s)}$. Their integration generates action
\eqref{mS} which is equivalent to the minimal half-integer superspin action.

The first order description of integer superspins ($s>1$) parallels the non-minimal half-integer superspin
description. There is a half-order invariant $\I_{\b\a(s-1)\ad(s-1)}$ \eqref{I3} and the action \eqref{S3}
is written with the help of two auxiliary superfields $\O_{\b\a(s-1)\ad(s-1)}$ and $\W_{\b\a(s-1)\ad(s-1)}$.
Their integration yields action \eqref{s3} which is equivalent to the known integer superspin action.

The hope is that the existence of these descriptions will simplify the investigation of
manifestly supersymmetric higher spin interactions. Moreover for non-supersymmetric higher spins, first
order formalism allows the study of higher spin dualities \cite{Casini:2003kf,Boulanger:2003vs}. Our
results may allow the examination of similar higher superspin dualities in superspace.


\begin{acknowledgments}
The work of K. K. is supported in part by the endowment of the Ford Foundation Professorship of Physics at
Brown University. Also K. K. gratefully acknowledges the support of the Brown Theoretical Physics Center.
\end{acknowledgments}

\bibliography{references}

\end{document}